\documentstyle[12pt]{article}
\begin{document}
\baselineskip=18pt
\pagestyle{empty}
\begin{center}
\bigskip

\rightline{CWRU-P16-99}

\vspace{0.5in}
{\Large \bf Geometry and Destiny}
\vspace{0.3in}

\vspace{.2in}
Lawrence M. Krauss$^1$ and Michael S. Turner$^{2,3}$\\
\vspace{.2in}
{\it $^1$Departments of Physics and Astronomy\\
Case Western Reserve University\\
Cleveland, OH~~44106-7079\\
e-mail:  krauss@theory1.phys.cwru.edu}\\
\vspace{0.1in}

{\it $^2$Departments of Astronomy \& Astrophysics and of Physics\\
Enrico Fermi Institute, The University of Chicago, Chicago, IL~~60637-1433\\
e-mail:  mturner@oddjob.uchicago.edu}\\

\vspace{0.1in}

{\it $^3$NASA/Fermilab Astrophysics Center\\
Fermi National Accelerator Laboratory, Batavia, IL~~60510-0500}\\
\vspace{0.2in}

({written for the {\it Gravity Research Foundation Essay Competition}})

\end{center}

\vspace{.3in}

\centerline{\bf SUMMARY}
\bigskip
\noindent The recognition that the cosmological constant may be non-zero
forces us to re-evaluate standard notions about the
connection between geometry and the fate of our Universe.
An open Universe can recollapse, and a closed Universe can expand forever.
As a corollary, we point out that there is no set of cosmological observations
we can perform that will unambiguously allow us to determine what the
ultimate destiny of the Universe will be.

\newpage
\pagestyle{plain}
\setcounter{page}{1}
\newpage

{}The traditional philosophy of General Relativity is that Geometry is
Destiny.   We teach undergraduates that the Universe can exist in one of three
different geometries, open, closed and flat, and that once we determine which
describes our Universe, this fixes its fate.

{}In the past few years, however, several features of conventional wisdom in
cosmology have fallen by the wayside.   By 1995 it was already clear that
fundamental observables, from the age of the Universe, to the baryon content,
and the nature of large-scale structure, all independently pointed to the
possible existence of a non-zero cosmological constant \cite{kraussturn}.
At the very least, there is now definitive evidence that matter, be it dark or
luminous, is not sufficiently abundant to result in a flat Universe today
\cite{omega_m}.  If we are to believe one of the generic
predictions of inflation---that we live in an almost exactly flat
Universe---a cosmological constant, or some form of energy very much like it is
the only way out.

{}This speculation received dramatic support a year ago, with independent
claims by two groups that Type 1a supernova, when used as standard candles,
indicated that the expansion of the Universe is {\it accelerating}
\cite{perl,kirsh}.  The simplest explanation of this result is the
presence of a cosmological constant.

{}Most recently, observations of the Doppler peak in the Cosmic Microwave
Background anisotropies have begun to provide more definitive evidence that we
live in a flat Universe today \cite{cbr}.  When this fact is combined with
the SN 1a data, and the data from large-scale clustering, a parameter
range of $\Omega_{M} \approx 0.3-0.4$ and $\Omega_{\Lambda} \approx
0.6-0.7$ appears to be strongly favored \cite{bestfit}.

{}While it is premature to claim, on the basis of the existing data, that a
$\Lambda$-dominated flat model actually describes our Universe, it is not
premature to explore its possible ramifications.  Recently, for example, an
analysis has been performed that suggests that this observation will have
important implications for the future of life in our Universe
\cite{kraussstark}. Here we focus on a more general feature
associated with the incorporation of a cosmological constant into our models: 
The one-to-one correspondence between geometry and evolution is forever lost.

{}The mathematical basis of this is described simply.  Einstein's equations
imply, for an isotropic and homogeneous Universe, the following evolution
equations for the cosmic scale factor, $R(t)$:
\begin{eqnarray}
 H^2\equiv { \left ( \dot R \over R \right )}^2 & = &
 {8 \pi G\over 3}  \rho_{\rm TOT}  - {k \over R^2}
\label{eq:rdot}  \\
{\ddot R  \over R} & = & -{4 \pi G\over 3}\sum_i\rho_i(1 + 3w_i)
\label{eq:rdotdot}
\end{eqnarray}
where $k$ is the signature of the 3-curvature,
the pressure in component $i$ is related to the energy
density by $p_i = w_i \rho_i$ and the total energy density
$\rho_{\rm TOT} = \sum_i\rho_i$.
The evolution of the energy density in component $i$ is determined by
\begin{equation}
{d\rho_i \over \rho_i} = -3(1+w_i){dR\over R}\ \
\Rightarrow \ \ \rho_i\ \propto\ R^{-3(1+w_i)}
\label{eq:firstlaw}
\end{equation}

{}All forms of normal matter satisfy the strong-energy condition,
$(\rho_i  + 3p_i)=\rho_i(1+3w_i)>0$, and so if
the Universe is comprised of normal matter,
the expansion of the Universe always decelerates, cf. Eq.~(\ref{eq:rdotdot}).
Also, since $\rho$ is positive for normal matter,
the first equation implies that $ \dot R /R$
remains positive and non-zero if $k \le 0$, and thus the
Universe expands forever.  Equation (\ref{eq:firstlaw}) and
the strong-energy condition imply that $\rho_i$ decreases more
rapidly than $R^{-2}$.  Thus, for $k>0$ there is necessarily
a turning point with $H=0$ and $\ddot R < 0$, and the Universe
must ultimately recollapse.  Geometry determines destiny.

{}However, a cosmological constant
violates the strong-energy condition, completely obviating the logic of the above
argument.   Recalling that  $p_\Lambda =-\rho_\Lambda$ for a
cosmological term, and that $p_M=0$
for matter, the above equations become,
\begin{eqnarray}
H^2 & = & {8 \pi G\over 3}  (\rho_M + \rho_{\Lambda})  - {k \over R^2}
\label{eq:rdot_lambda} \\
{\ddot R \over R} & = & -{4 \pi G\over 3}(\rho_M -2 \rho_{\Lambda})
\label{eq:rdotdot_lambda}
\end{eqnarray}

{}Since $\rho_{\Lambda} = $ constant, while $\rho_M \propto R^{-3}$,
even if $k >0$, as long as $H > 0 $ when $\rho_{\Lambda}$
comes to dominate the expansion, it will remain positive forever,
and as is well known, the expansion will ultimately accelerate,
$R(t) \rightarrow e^{Ht}$ with $H=\sqrt{8\pi G\rho_\Lambda /3}$.

{}One conventionally defines the scaled energy density
$\Omega \equiv \rho_{\rm TOT} /\rho_{\rm crit} =
8 \pi G \rho /3H^2$, so that $\Omega -1 = k/H^2R^2$.
Thus the sign of $k$ is determined by whether $\Omega$ is
greater than or less than 1.
In this way, a measurement of $\Omega$ at any epoch -- including
the present -- determines the geometry of the Universe.  However,
we can no longer claim that the magnitude of $\Omega$ uniquely
determines the fate of the Universe.

{}This decoupling between $\Omega$ and destiny can also be seen using Sandage's
deceleration parameter
$q\equiv -(\ddot R/R)/H^2$, which,
by using Eqs.~(\ref{eq:rdot},\ref{eq:rdotdot}), can be written as
\begin{equation}
q = {\Omega\over 2} + {3\over 2}\sum_i w_i\Omega_i
\end{equation}
The sign of $q$, and thus the deceleration of the
Universe at any given epoch depends upon the equation of state
and not on $\Omega$ alone.

{}While in the presence of a cosmological constant, $\Omega$ no longer determines
the ultimate fate of the Universe, it is useful in determining how
small a cosmological constant could be at the present time and still stop the
eventual collapse of a closed Universe.  For a closed, matter-only Universe,
the scale factor at turnaround is
\begin{equation}
{R/R_0} = {\Omega_0 \over \Omega_0 - 1}
\end{equation}
While all the evidence today suggests
that $\Omega_0 \le 1$, existing uncertainties could allow $\Omega_0$ to
be as large say as 1.1.  For $\Omega_0=1.1$ the scale
factor at turnaround is $11R_0$.   Since the density of matter decreases as
$R^{-3}$, this means that an energy density in a cosmological term as
small as $1/1000th$ the present matter density will come to dominate the
expansion before turnaround and prevent forever recollapse.
A cosmological constant this small,
corresponding to $\Omega_\Lambda \sim 0.001$, is completely
undetectable by present, or foreseeable observational probes.

{}Alternatively, it may seem that if we can unambiguously determine that
$k <0$ then we are assured the Universe will expand forever.   However, this is
the case only as long as the cosmological constant is positive.  Since we have
no theory for a cosmological constant \cite{weinberg},
there is no reason to suppose that this
must be the case.   When the cosmological constant is negative, the energy
density associated with the vacuum is constant and {\it negative}.   In this
case, from Eqs.~(\ref{eq:rdot_lambda},\ref{eq:rdotdot_lambda}),
one can see that not only is the ultimate expansion
guaranteed to decelerate, but recollapse is also inevitable,
{\it no matter how small the absolute value of $\Omega_\Lambda$ is}.

{}Finally, what if we indeed ultimately verify a non-zero cosmological
constant at the present
time, as current observations suggest?  Are we not then guaranteed an eternal
expansion?  The answer is again no.   As is well known, we have no guarantees
that what we observe to behave as a cosmological constant is in fact the
actual ground-state vacuum-energy density of the Universe.  Any scalar field
which is not at the minimum of its potential will, as long as the age of the
Universe is small compared to the characteristic time it takes for the field to
evolve in its potential, mimic a cosmological term in Einstein's equations. 
Until the field evolves to its ultimate minimum, we cannot derive the
asymptotic solution of these equations in order to determine our destiny.

{}We thus arrive at the following set of possibilities.  As this
classification makes clear, it is the ultimate equation
of state, not geometry, that determines the fate of the Universe.  The key
consideration is the value of $\ddot R$ at any potential turning point
(i.e., where $H=0$).  If it is negative, which requires $\sum_i\rho_i(
1+3w_i)>0$, recollapse occurs; otherwise expansion resumes and continues
eternally.

\vskip 0.15in
\noindent{
{\bf RECOLLAPSE}}
\vskip 0.1in

1.  Closed ($k > 0$) Universe: $\rho_M > 2\rho_{\Lambda}$ when $H=0$

2.  Open,flat ($k \le 0$), or closed ($k > 0$) Universe: $\rho_{\Lambda} <
0$

\vskip 0.1in
\noindent{
{\bf ETERNAL EXPANSION}}
\vskip 0.1in

1.  Closed ($k > 0$) Universe : $\rho_M < 2\rho_{\Lambda}$ before $H=0$

2.  Open or flat ($k \le 0$) Universe: $\rho_{\Lambda} \ge 0$

\vskip 0.1in

{}For the simplest possibility,
a Universe with matter and positive cosmological constant,
the dividing point between expansion forever and recollapse
can be expressed simply:  Eternal expansion is
inevitable if and only if \cite{carrolletal}
\begin{equation}
\Omega_\Lambda > 4\Omega_M \left\{
\cos \left[ {1\over 3}\cos^{-1}(\Omega_M^{-1} -1)
+ {4\pi \over 3}\right] \right\}^3 .
\end{equation}
Given Einstein's association with the cosmological constant,
we would be quite remiss in not mentioning the intermediate case,
his static Universe.  A static, but unstable, cosmological
solution obtains for
\begin{equation}
\rho_M=2\rho_\Lambda \ \ \ {\rm and}\ \ \  R
= k^{1/2}/\sqrt{8\pi G\rho_{\Lambda}}
\end{equation}
The above classification can of course be generalized to any other
form of energy that violates the strong-energy condition and/or mimics a
cosmological constant. In such cases, the equation of state will generally
vary with time.

{}Indeed, because the equation of state of the Universe can change, we
may never be confident that any presently inferred dynamical evolution can be
extrapolated indefinitely into the future.  Put another way: even if the
presently inferred cosmological constant turns out to be a red herring, we
cannot definitively argue that a closed Universe will recollapse or that an
open Universe will expand forever. A smaller, presently unobservable value of
$\Lambda$ could always alter the ultimate fate of the Universe. 
In a true sense therefore, perhaps only knowledge of a fundamental theory of
everything, one that predetermines both the asymptotic values of both $\Omega$
and $\Lambda$, will allow us ultimate knowledge of the ultimate state of the
Universe.  If instead, the fundamental parameters in
our observed Universe arise from a probability distribution based on some
underlying theory, then the future is ultimately unknowable.

{}While these features of the Universe
have been implicit in Einstein's equations
since they were first written down, they tended to disappear from the popular
lore shortly after the cosmological constant did.  It is therefore important,
now that we appear to be living in a Universe with non-zero cosmological
constant, to re-acquaint ourselves with their implications.

{}Before concluding, we note one additional complication brought on by
possible quantum field theoretic phenomena.
While the 4-geometry of the observable Universe is
classically invariant, its 3-geometry depends upon the choice of the
constant-time hypersurfaces.   It is distinctly possible that a
change in equation-of-state will also change the natural choice of
hypersurfaces, and thus alter the inferred 3-geometry.  A well known example
involves the nucleation of a bubble of true vacuum in the midst of a background
of false vacuum \cite{guthwein}.   Observers inside
this bubble will infer an
open geometry, while those outside the bubble will observe the bubble to
collapse into a black hole.  In the different cases an infinite spatial region
is interchanged with an infinite temporal one.  Further, destiny could
also change, as the bubble could have nucleated within a Universe
destined for recollapse.  Of course, under the conditions
of the change in the equation of state defined by the situation described
above, the future will be truly unknowable for another reason. It is unlikely
that any life-form that evolved in one vacuum would survive the transition into
the other.

{}Returning to our central thrust, the result
that geometry and destiny are decoupled is in one sense disappointing.  The
hope that we could, via a finite set of cosmological observations that might be
completed within the next decade, determine eternity was very satisfying.  
Nevertheless, what we lose in predictive power we gain in fundamental
excitement.   The microphysics that might generate a non-zero cosmological
constant or a scalar field that mimics one will no doubt be central to much of
the forefront theoretical and experimental research in the next century, if not
the next millennium.   The new uncertainty in our ultimate destiny thus opens an
exciting door that may lead to a deeper understanding of our ultimate origin.

\section*{Acknowledgments}
This work was supported in part by the DOE (at Chicago,
Fermilab and Case Western Reserve) and by
the NASA (at Fermilab through grant NAG 5-7092).


\begin{thebibliography}  {gcages}

\bibitem{kraussturn} L.M.~Krauss and M.S.~Turner, {\it Gen. Rel. Grav.}
{\bf 27}, 1135 (1995); J.P.~Ostriker and P.J.~Steinhardt, {\it Nature}
{\bf 377}, 600 (1995).

\bibitem{omega_m} L.M.~Krauss "The New Cosmology and Dark Matter", to appear in
{\it Proceedings of Sheffield International Workshop on Dark Matter,
Sept 1998} (hep-ph/9807376); N.A. Bahcall and X. Fan,
{\it Proc. Nat. Acad. Sci.} {\bf 95}, 5956 (1998);
M.S. Turner, {\it Physica Scripta}, in press (1999) (astro-ph/9901109).

\bibitem{perl}
S. Perlmutter et al, {\it Astrophys. J.}, in press (1999) (astro-ph/9812133).

\bibitem{kirsh}
A.G.~Riess et al, {\it Astron. J.} {\bf 114}, 722 (1998).

\bibitem{cbr} See e.g., K. Coble et al, {\it Astrophys. J.}, submitted
(1999) (astro-ph/9902195); C. Lineweaver, {\it Astrophys. J.} {\bf 505}, 69
(1998).

\bibitem{bestfit} S. Perlmutter, M.S. Turner, and M. White,
Phys. Rev. Lett., submitted (1999) (astro-ph/9901052); L.M.~Krauss,
{\it Astrophys. J.} {\bf 501}, 461 (1998).

\bibitem{kraussstark} L.M.~Krauss and G.D.~Starkman, {\it Astrophys. J.},
submitted (1999) (astro-ph/9902195).

\bibitem{weinberg}
S. Weinberg, {\it Rev. Mod. Phys.} {\bf 61}, 1 (1989).

\bibitem{carrolletal}
S.M.~Carroll, W.H.~Press and E.L.~Turner,
  {\it Ann. Rev. Astron. \& Astrophys.} {\bf 30}, 499 (1992).

\bibitem{guthwein} A. Guth and E.J. Weinberg, {\it Nucl. Phys. B}
{\bf 212}, 321 (1983).

\end{thebibliography}
\end{document}